\documentclass[fleqn,usenatbib]{mnras}
\usepackage{graphicx}
\usepackage{amssymb}
\usepackage{amsmath}
\usepackage{mathabx}
\usepackage{txfonts}
\usepackage[T1]{fontenc}
\usepackage{ae,aecompl}
\usepackage{cleveref}

\usepackage{pdfsync}
\graphicspath{{figs/}}

\usepackage[dvipsnames]{xcolor}

\title[Dynamical friction and thermal effects]{Dynamical friction with radiative feedback -- II. High resolution study of the subsonic regime}

\author[D. A. Velasco Romero et F. Masset.]{%
David A. Velasco Romero$^{1,2,3}$\thanks{E-mail: david.velasco@icf.unam.mx} and Fr\'ed\'eric S. Masset$^{2}$\\
$^{1}$Universidad Aut\'onoma del Estado de Morelos, Av. Universidad s/n, 62210 Cuernavaca, Mor., Mexico\\
$^{2}$Instituto de Ciencias F\'isicas, Universidad Nacional Aut\'onoma de M\'exico, Av. Universidad s/n, 62210 Cuernavaca, Mor., Mexico\\
$^{3}$Institute for Computational Science, University of Zurich
}

\date{Accepted XXX. Received YYY; in original form ZZZ}

\pubyear{2020}

\begin{document}
\label{firstpage}
\pagerange{\pageref{firstpage}--\pageref{lastpage}}
\maketitle

\begin{abstract}
  Recent work has suggested that the net gravitational force acting on a massive and luminous perturber travelling through a gaseous and opaque medium can have same direction as the perturber's motion (an effect sometimes called negative dynamical friction). Analytic results were obtained using a linear analysis and were later confirmed by means of non-linear numerical simulations which did not resolve the flow within the Bondi sphere of the perturber, hence effectively restricted to weakly perturbed regions of the flow (paper~I). Here we present high resolution simulations, using either 3D Cartesian or 2D cylindrical meshes that resolve the flow within the Bondi sphere. We perform a systematic study of the force as a function of the perturber's mass and luminosity, in the subsonic regime.  We find that perturbers with mass $M$ smaller than a few $M_c\sim \chi c_s/G$ are subjected to a thermal force with a magnitude in good agreement with linear theory ($\chi$ being the thermal diffusivity of the medium, $c_s$ the adiabatic sound speed and $G$ the gravitational constant), while for larger masses, the thermal forces are only a fraction of the linear estimate that decays as $M^{-1}$. Our analysis confirms the possibility of negative friction (hence a propulsion) on sufficiently luminous, low-mass embryos embedded in protoplanetary discs.  Finally, we give an approximate expression of the total force at low Mach number, valid both for sub-critical ($M<M_c$) and super-critical ($M>M_c$) perturbers.

\end{abstract}

\begin{keywords}
hydrodynamics -- gravitation -- planet-disc interactions -- accretion, accretion discs -- black hole physics
\end{keywords}

\defcitealias{2017MNRAS.465.3175M}{MV17}
\defcitealias{2019MNRAS.483.4383V}{paper~I}


\section{Introduction}
\label{sec:introduction}

As studied analytically by \citet{1999ApJ...513..252O} and confirmed by the numerical simulations of \citet{1999ApJ...522L..35S}, a point-like massive perturber travelling through an adiabatic, initially homogeneous, gaseous medium is subjected to a gravitational drag, or dynamical friction. Relaxing the assumption of adiabaticity (i.e., allowing heat to diffuse through the gas), one finds an enhanced drag on the perturber \citep[][hereafter paper~I]{2019MNRAS.483.4383V}.  If, furthermore, the perturber is luminous and heats its surroundings at the rate $L$, a new component of the gravitational force appears, dubbed in previous work heating force, and directed along the motion of the perturber.  When the luminosity is sufficiently large, the net force can be a propulsion on the perturber \citep[][hereafter MV17]{2017MNRAS.465.3175M}, which can have important implications for scenarios of planet formation \citep{2015Natur.520...63B,2017MNRAS.465.3175M,2017arXiv170401931E}.  In the limit of a low Mach number, \citetalias{2019MNRAS.483.4383V} find that the total force exerted over the perturber is the sum of three forces: (i) the drag force that would be exerted if the gas were adiabatic, (ii) the thermal cold force that corresponds to an additional drag when thermal diffusion occurs (the perturber surroundings are then colder than they would if the gas were adiabatic, and are flooded with additional material, mainly trailing), and (iii) the heating force if the perturber is luminous. These results were obtained using non-linear numerical simulations with nested meshes which did not resolve the Bondi sphere, thereby restricted to weakly perturbed parts of the flow. 
While supersonic motion has drawn a lot of attention in recent work \citep[e.g.][]{2011MNRAS.418.1238C,2016A&A...589A..10T}, here we restrict ourselves to subsonic motion, which has a considerable interest when radiative feedback is taken into account.
The aim of this work is to revisit, in the limit of a low Mach number, the behaviour of the total force and of its thermal components (i.e. the thermal cold force and the heating force) when the flow within the Bondi sphere is resolved. We make use of the hydrocode FARGO3D\footnote{\texttt{https://bitbucket.org/fargo3d/public}} with a nested meshes capability \citepalias{2019MNRAS.483.4383V} to perform our simulations. In section~\ref{sec:problem-description} we describe the system at study, and we provide a summary of the analytic results of \citetalias{2017MNRAS.465.3175M} and their confirmation through numerical simulations in \citetalias{2019MNRAS.483.4383V}. In section~\ref{sec:numerical_implementation} we present a description of our implementation of 2D cylindrical meshes, introduce the implementation of a more realistic thermal diffusivity, needed for strongly perturbed parts of the flow, and present an implementation for the release of heat into the ambient medium. In section~\ref{sec:results} we present our results. We first evaluate the magnitude of the net force, then turn to the yield of the heating force. We subsequently analyse how thermal forces differ from the analytical estimate arising from a linear analysis, and find a cut off at larger mass. We finally analyse the properties of the flow at the Bondi scale around a massive perturber using a passive tracer, for different luminosities of the perturber. In section~\ref{sec:simple-cut-model} we present a simple yet effective model to account for the behaviour of thermal forces at larger masses and in section~\ref{sec:disc-concl} we summarise our findings and give our conclusions.

\section{Description of the problem.}
\label{sec:problem-description}
The system at study consists of a point-like perturber of mass $M$ and luminosity $L$ immersed in an initially uniform gas medium, and having a constant velocity. It is naturally simpler to perform a change of frame and consider the perturber to be fixed at the origin of a Cartesian reference frame, whereas the gas has initially a uniform advection velocity $V$ directed along the $z$~axis. The thermal diffusion in the gas follows Fourier's law for heat transfer with a thermal diffusivity $\chi$.

\subsection{Governing equations}
\label{sec:governing-equations}
For consistency with \citetalias{2017MNRAS.465.3175M} and \citetalias{2019MNRAS.483.4383V} we use  $\rho$, $\boldsymbol{v}$ and $e$ to respectively denote the gas density, velocity and density of internal energy, and we use $\rho_0$, $\boldsymbol{v}_0$ and $e_0$  to denote their respective unperturbed quantities. The governing equations of our system are then: 

\begin{eqnarray}
\label{eq:1}
\partial_t \rho + \nabla \cdot (\rho\boldsymbol{v}) = 0\\
\label{eq:2}
\partial_t (\rho\boldsymbol{v}) + \nabla\cdot\left(\rho\boldsymbol{v}\otimes \boldsymbol{v}+p\right)=  -\rho\nabla\Phi \\
\label{eq:3}
\partial_t e + \nabla \cdot (e\boldsymbol{v}) + p\nabla \cdot \boldsymbol{v} + \nabla \cdot \left( \chi\rho\nabla \frac{e}{\rho} \right) = L \delta (\boldsymbol{r})
\end{eqnarray} 

where $\Phi= -GM/|\boldsymbol{r}|$ represents the gravitational potential of the perturber, the last term of the left hand side (L.H.S.) of Eq.~\eqref{eq:3} the divergence of the  energy flux coming from heat diffusion and obtained from Fourier's law. Our system is closed by the equation of state $p=(\gamma-1)e$ where $p$ represents the pressure and $\gamma$ the adiabatic index.

\subsection{Summary of previous results}
\label{sec:linear-study-summary}

Here we recapitulate the results of the analytic study using linear perturbation theory for the heating force \citepalias{2017MNRAS.465.3175M} and for the thermal cold force \citepalias{2019MNRAS.483.4383V}. Analytic results were obtained in the limit of low and high Mach numbers. The scope of the present work being the subsonic regime, we summarise only the results obtained in the limit of a low Mach number.

The thermal cold force, i.e. the additional contribution to the net force arisen from thermal diffusion, is given by Eq.~(48) of \citetalias{2017MNRAS.465.3175M}:
\begin{equation}
  \label{eq:4}
  F^\mathrm{cold}_\textrm{thermal}=-\frac{2\pi G^2M^2\rho_0(\gamma-1)}{c^2_s}\text{sign}(V).
\end{equation}
We emphasise its independence on the values of the thermal diffusivity $\chi$ and velocity, and that it is directed against the motion of the perturber, as it is an additional drag force. We proceed with the contribution to the net force coming from the gravitational feedback of the medium onto a luminous perturber \citepalias{2017MNRAS.465.3175M}:
\begin{equation}
  \label{eq:5}
F_\mathrm{heating} = \frac{\gamma(\gamma-1)GML}{2\chi c_s^2}\text{sign}(V).
\end{equation}
This ``heating'' force is directed along the motion of the perturber, promoting therefore its propulsion over the ambient medium. Eqs.~\eqref{eq:4} and~\eqref{eq:5} represent the asymptotic values of  the forces at vanishing Mach number. From these expressions we can find the critical luminosity $L_c$ at which these forces cancel out, which reads \citep{2017MNRAS.472.4204M,2019MNRAS.483.4383V}:
\begin{equation}
  \label{eq:6}
	L_c = \frac{4\pi G M \rho_0 \chi}{\gamma}.
\end{equation}
In the case of a small Mach number, a perturber with a luminosity $L=L_c$ is therefore subjected to the adiabatic drag force of \citet{1999ApJ...513..252O}. As done in  \citetalias{2019MNRAS.483.4383V}, we recast our forces in terms of the following force:
\begin{equation}
  \label{eq:7}
  F_0=\frac{4\pi(GM)^2\rho_0}{c_s^2}, 
\end{equation}
with which the cold thermal force reads: 
\begin{equation}
  \label{eq:8}
  F_\mathrm{thermal}^\mathrm{cold}=-\frac{\gamma-1}{2}F_0, 
\end{equation}
the heating forces reads: 
\begin{equation}
  \label{eq:9}
  F_\mathrm{heating}=\frac{\gamma-1}{2}\frac{L}{L_c}F_0,
\end{equation}
and the adiabatic force in the subsonic regime reads:
\begin{equation}
  \label{eq:10}
  F_\mathrm{adi}=-F_0{\cal M}^{-2}I_\mathrm{subsonic} 
\end{equation}
where $I_\mathrm{subsonic}$ is given by Eq.~(14) of \citet{1999ApJ...513..252O}, while in the limit of  small Mach number, $I_\mathrm{subsonic} \to{\cal M}^3/3$, so that:
\begin{equation}
  \label{eq:11}
  F_\mathrm{adi} \to -\frac{\cal M}{3} F_0.
\end{equation}
In the limit of a low Mach number and using the linearised equations of the perturbed flow, we show in \citetalias{2019MNRAS.483.4383V} that the net force experienced by a massive and luminous perturber in a thermally diffusive gas medium results is the sum of these three forces:
\begin{equation}
  \label{eq:12}
  F_\mathrm{total}= F_\mathrm{adi} +  F_\mathrm{thermal}^\mathrm{cold} + F_\mathrm{heating}.
 \end{equation}
The linearisation of the equations for the perturbed flow is valid outside the Bondi radius defined as:
\begin{equation}
  \label{eq:13}
R_B=\frac{GM}{c^2_s},
\end{equation}
inside which the flow is highly non-linear. As done in \citetalias{2017MNRAS.465.3175M}, we define a critical mass:
\begin{equation}
  \label{eq:14}
  M_c=\frac{c_s \chi}{G}
\end{equation}
under which the heat diffusion time within the Bondi radius is  shorter than the acoustic time $R_B/c_s$, and vice-versa. The characteristic length for the advection-diffusion problem is given by the cutoff distance:
\begin{equation}
  \label{eq:15}
\lambda=\frac{\chi}{\gamma V},
\end{equation}
and the time that it takes the hot plume of characteristic size $\lambda$ to settle is given by:
\begin{equation}
  \label{eq:16}
\tau=\frac{\lambda^2}{\chi}.
\end{equation}

\section{Numerical Implementation.}
\label{sec:numerical_implementation}
\subsection{Cylindrical and Cartesian meshes}
As done in \citetalias{2019MNRAS.483.4383V}, we initially performed simulations with 3-dimensional nested Cartesian meshes. The main domain is given by $x_{min}=0$, $x_{max}=l$, $y_{min}=0$, $y_{max}=l$, $z_{min}=-l$ and $z_{max}=l$ (with $l$ being half the box length), each nested mesh covering half the domain in each coordinate of the immediately coarser mesh while doubling the resolution (so that all levels have the same number of zones). Given the high resolution required to adequately describe the dynamics inside the Bondi radius, we also resorted to the rotational symmetry of the problem at hand to lower the dimension of our test bed by performing 2D cylindrical simulations similar to those of \citet{Kim2009}. These new 2D cylindrical simulations have a main domain with $z_{min}=-l$, $z_{max}=l$, $r_{min}=0$ and $r_{max}=l$. We also take advantage of nested meshes to increase the resolution around the perturber with this 2D setup, with lower increments in computational cost (both in execution time and memory usage) given the reduced dimensionality of the setup.

We use an adiabatic index $\gamma=1.4$. The perturber is located at $\boldsymbol{r}_0$ and its potential follows Plummer's law:

\begin{equation}
  \Phi(\boldsymbol{r}_{i,j,k})=-\frac{GM}{\sqrt[]{|\boldsymbol{r}_{i,j,k}-\boldsymbol{r}_0|^2+\epsilon^2},}
\end{equation}
where $\epsilon$ is the smoothing length. In this work we use $\epsilon=0.2R_B$. If one considers the smoothing length as a proxy for the perturber's physical radius, the ratio $\epsilon/R_B$ is approximately that of a Mars-sized embryo at $1$~au of a solar mass star, in a disc with aspect ratio $h=0.05$.

We monitor at each time step of the simulation the $z$-component of the force exerted by the system over the perturber:
\begin{equation}
F= \sum_l \frac{z_lGM\rho_lV_l}{(r'^2_l+\epsilon^2)^{3/2}},
\end{equation}
where $r'_l$ is the distance from the centre of the $l$th cell to the position of the perturber.

In order to measure the heating force exerted onto the perturber in a steady state, we need to reach a physical time of several times $\tau$. We chose to go up to $t_\text{end}=100\tau$ for masses at which $\tau$ is at least an order of magnitude larger than the acoustic time over the Bondi radius ($R_B/c_s$). For larger masses we increased the physical time to be reached as $t_\text{end}=100\tau M/M_c$, in order to allow the force to settle. This requirement on the time of the simulation imposes a minimal size for the main domain (i.e. the lowest resolution mesh) of the simulation: in order to avoid the acoustic sphere triggered by the perturber at $t=0$ to reach and bounce off the boundaries we need a box half length of at least $l=t_\text{end}(c_s+V)$. Describing the dynamics inside the Bondi radius brings upon the need to solve this radius with a high number of cells (from $8$ up to $64$ cells in our simulations). Reaching this requirement implies a higher computational toll as we decrease the mass (and therefore $R_B$).

\subsection{Slow initial growth of the perturber's mass}
\label{sec:taper-mass-depos}
In the initial stages of our parameter exploration we came upon the early formation of vortices in the adiabatic scenario ($\chi=0$), as found for a subsonic perturber by \citet{Kim2009}. We found that including a taper on the mass deposition delays the formation of such vortices, which allows a clean comparison against the diffusive scenario where we have not observed this phenomenon. In this study we use a mass taper time of several times $R_B/c_s$ (from $0.5R_B/c_s$ to $25R_B/c_s$). We did not search for the maximum rate of mass deposition that prevents the appearance of vortices. We found that as long as the vortices do not appear, the force measured in our adiabatic simulations is within a few percent of the analytic expression of \citet{1999ApJ...513..252O}, so when subtracting the adiabatic force from the force measured in any of our runs, we make use of that analytical expression.
 
\subsection{Thermal diffusion}
\label{sec:thermal-diffusion}

As in \citetalias{2019MNRAS.483.4383V} we were dealing with weakly perturbed flow, the thermal diffusivity in the gas was nearly constant. For the present work, we have implemented a more realistic model as found for stellar media \citep{1958RA......5..204S} or for circumstellar discs \citep{1996ApJ...461..933K}, with the thermal diffusion coefficient as a function of the temperature and density:

\begin{equation}
  \label{eq:17}
    \chi = \chi_0\left(\frac{T}{T_0}\right)^3\left(\frac{\rho}{\rho_0}\right)^{-2}.
  \end{equation}
  For a gas with density and temperature comparable to those of protoplanetary discs at planet forming distances from the star, this expression stems from the following expression of the thermal diffusivity \citep{pbk11}:
  \begin{equation}
    \label{eq:18}
    \chi = \frac{16\gamma(\gamma-1)\sigma T^4}{3\kappa\rho^2H^2\Omega^2},
  \end{equation}
  where $\sigma$ is Stefan's constant, $\kappa$ the opacity, $H$ the disc's pressure scale height and $\Omega$ the local orbital frequency. Eq.~\eqref{eq:17} disregards any variation of the opacity as a function of $\rho$ and $T$.
  
For our numerical implementation we compute the thermal diffusivity at each cell interface (where heat fluxes are evaluated) as
\begin{equation}
    \chi_{i-\frac12}=\chi_0\left(\frac{e_i+e_{i-1}}{2e_0}\right)^3\left(\frac{2\rho_0}{\rho_i+\rho_{i-1}}\right)^5 = 4\chi_0 \frac{(e_i+e_{i-1})^3}{(\rho_i+\rho_{i-1})^5}.
\end{equation}
Then, similarly to Eqs.~(19) to~(21) of \citetalias{2019MNRAS.483.4383V}, the energy fluxes multiplied by their corresponding surfaces are:
\begin{eqnarray}
  f^x_{i-\frac12} &=& \chi_{i-\frac12}\left(\frac{e_i}{\rho_i}-\frac{e_{i-1}}{\rho_{i-1}}\right)\frac{S_x}{\Delta x}\\
  f^y_{j-\frac12} &=& \chi_{j-\frac12}\left(\frac{e_j}{\rho_j}-\frac{e_{j-1}}{\rho_{j-1}}\right)\frac{S_y}{\Delta y}\\
  f^z_{k-\frac12} &=& \chi_{k-\frac12}\left(\frac{e_k}{\rho_k}-\frac{e_{k-1}}{\rho_{k-1}}\right)\frac{S_z}{\Delta z}.
\end{eqnarray}
The density of internal energy is then updated as prescribed by Eq.~(22) of \citetalias{2019MNRAS.483.4383V}.

\subsection{Heat release}
\label{sec:heat-release-1}
In \citetalias{2019MNRAS.483.4383V} we used a simple prescription to update the internal energy of the cells near the perturber. It consisted of a uniform release of the planet's luminous energy within the eight cells nearest to the planet's position $r_0$, as described by \citet{2015Natur.520...63B}. Our resolution in the present work is so high that in many runs we marginally resolve what would be the photons' mean free path in our highest resolution nested mesh.  We therefore need to resort to a more elaborated implementation of the heat release by the perturber, and use the following prescription:
\begin{equation}
    e_{i,j,k}^{n+1} = e_{i,j,k}^{n} + \frac{L}{8\pi \epsilon^3} \exp{\left(-\frac{r'_{i,j,k}}{\epsilon}\right)}\Delta t,
\end{equation}
where $r'_{i,j,k}=|\boldsymbol{r}_{i,j,k}-\boldsymbol{r}_0|$ is the distance from the cell's centre to the perturber's position, and $\epsilon=0.2R_B$ is the same smoothing length of the Plummer's potential.  In doing so, we assume that $\epsilon$ has same order of magnitude than the mean free path of the photons coming from the perturber. In the context of embryos embedded in protoplanetary discs, such assumption is reasonable. Tab.~\ref{tab:1} lists the parameters that can be derived from the fiducial cases studied by \citetalias{2017MNRAS.465.3175M}. The mean free path of photons $\bar{\ell}$ is obtained from:
\begin{equation}
  \label{eq:31}
  \bar{\ell}=\frac{1}{\rho_0\kappa}.
\end{equation}
The last line of that table is the ratio of the Bondi radius of a perturber of critical mass to the mean free path of photons. It appears to be of same order of magnitude than the ratio of the release length to the Bondi radius used in the present study ($\epsilon/R_B=0.2$).

\begin{table}
    \centering 
    \begin{tabular}{l|c|c}
        \hline 
        Parameter & Value at $t=300\text{kyr}$ &  Value at $t=1\text{Myr}$\\
        \hline 
        $\kappa$ $(\text{cm}^2/\text{g})$ & $0.92$  &  $0.497$ \\
        $M_c$ $(\text{g})$ & $2.55 \times 10^{28}$  &  $8.81 \times 10^{26}$ \\
        $\bar{\ell}$ $(\text{cm})$ & $2.12 \times 10^{10}$  &  $0.86 \times 10^{10}$ \\
        $R_{B,M_c}$ $(\text{cm})$ & $17 \times 10^{10}$  &  $1.63 \times 10^{10}$ \\
        $\bar{\ell}/R_{B,M_c}$ & $0.125$ & $0.52$ \\
        \hline 
    \end{tabular}
    \caption{\label{tab:1}Parameters derived from table~1 of \citetalias{2017MNRAS.465.3175M} for the disc models of \citet{2015A&A...575A..28B} at $r = 3\text{au}$. The parameters of the first column are successively the opacity, the critical mass, the photon's mean free path, Bondi's radius for a critical mass and the ratio of the two previous rows. The first four parameters have a lower value at $t=1\text{~Myr}$ than at $t=300\text{~kyr}$. The drop of Bondi's radius for the critical mass is smaller than that of the critical mass itself, as the sound speed is smaller at larger time (see table~1 of \citetalias{2017MNRAS.465.3175M}).}
\end{table}{}

\subsection{Tracing the gas}
\label{sec:tracing-gas}
In order to track the gas we use a passive tracing field $\Psi\equiv z_0$ to ``label'' each fluid element with its initial vertical position. This field obeys the following advection equation: 
\begin{equation}
  \label{eq:19}
  \partial_t \Psi + \boldsymbol{v} \cdot \nabla \Psi = 0,
\end{equation}
which differs from the evolution equation of conservative quantities as the second term is not $\nabla\cdot(\Psi\boldsymbol{v})$. This renders the passive tracer insensitive to compression or expansion, and as such it can be thought of, for 2D flows, as an ``optically thick'' tracer, as opposed to an ``optically thin'' tracer, such as the density, that would obey the standard continuity equation.
We use the intermediate field  $\Psi'=\rho\Psi$, which is evolved according to:
\begin{equation}
  \label{eq:20}
      \partial_t \Psi' +   \nabla \cdot \left(\Psi' \boldsymbol{v}\right) = 0,
    \end{equation}
    and can therefore be updated using the same transport routines as the density, as it obeys a similar equation.
Expanding this equation in terms of $\rho$ and $\Psi$, then subtracting Eq.~\eqref{eq:1}, we see that $\Psi$
obeys Eq.~\eqref{eq:19}.
We can then keep track of the evolution of the field $\Psi$ by evolving the field $\Psi'$ along with the other conservative fields. We retrieve the value of $\Psi$ upon a conservative update from the new values of $\Psi'$ and $\rho$ using:
\begin{equation}
    \Psi^{n}_{i,j,k} = \frac{\Psi^{'n}_{i,j,k}}{\rho^{n}_{i,j,k}}.
\end{equation}

\section{Results}
\label{sec:results}
\subsection{Net force}
\label{sec:net-force}
\begin{figure*}
\includegraphics[width=\textwidth]{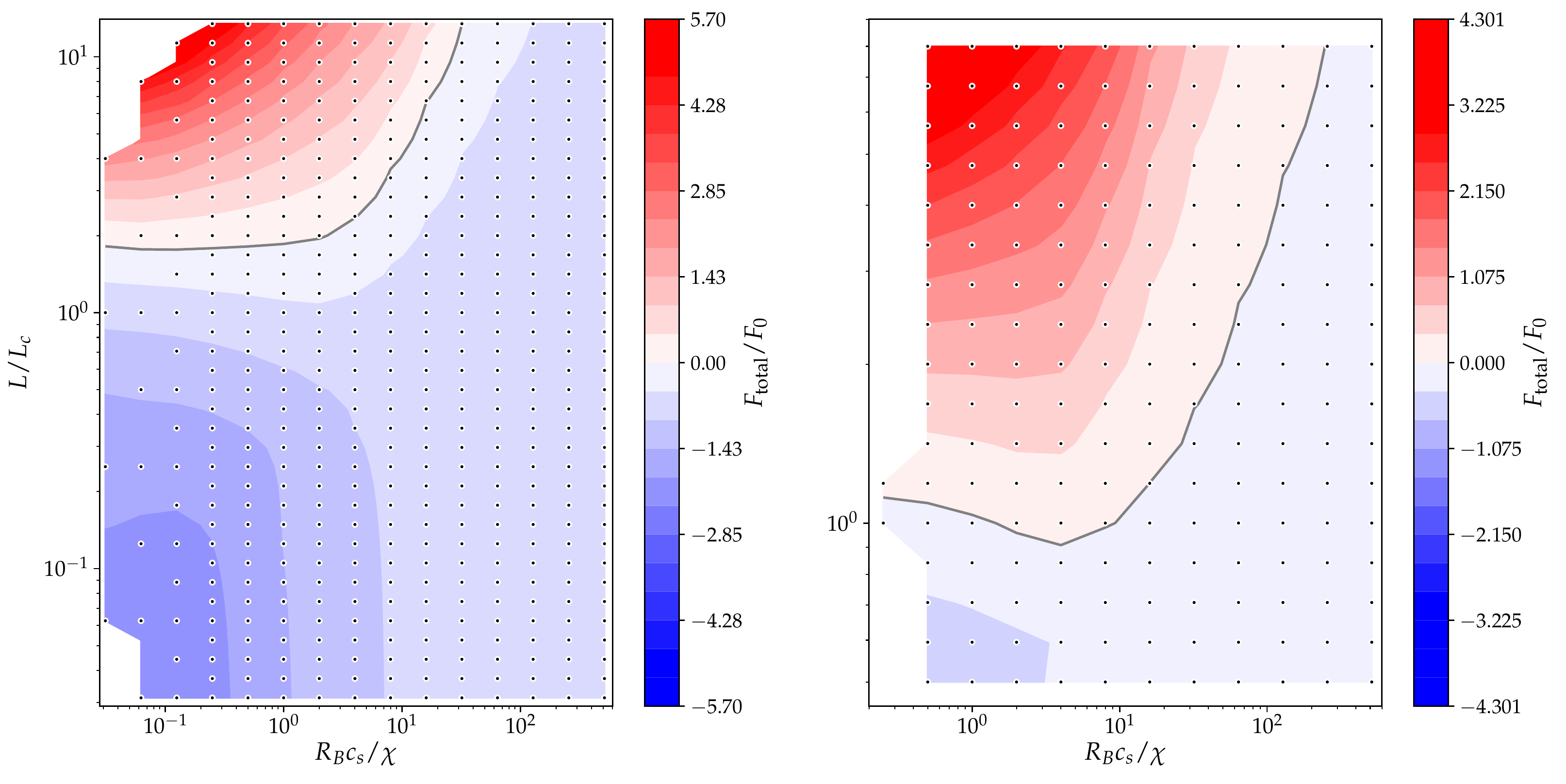}
\caption{Colour maps of the net force. The left plot shows the results for ${\cal M} = 0.4$ while the right plot shows the results for ${\cal M} = 0.1$. The red region (in the electronic version of this manuscript) corresponds to a positive net force (propulsion) while the blue region (in the electronic version of this manuscript) corresponds to negative net force (drag). We observe that for super-critical masses ($M/M_c=R_Bc_s/\chi> 1$) the minimum luminosity required to get a net propulsion scales roughly as $R_B$ (or as the mass).}
\label{fig:1}
\end{figure*}

We present in Fig.~\ref{fig:1} a study of the net force for two values of the Mach number in the subsonic regime (${\cal M}=0.4,0.1$). We performed simulations for a 2D array of parameters $(M,L)$, the mass $M$ being in geometric sequence around the critical mass $M_c$, and the luminosity $L$ spanning a geometric sequence of around the critical luminosity $L_c$ of the corresponding mass [as $L_c$ depends on $M$, see Eq.~\eqref{eq:6}]. We normalise the force measured in the simulations to the force $F_0$ given by Eq.~\eqref{eq:5}.  The vertical axis corresponds to the luminosity ratio $L/L_c$ whereas the horizontal axis corresponds to the mass ratio $M/M_c\equiv R_B c_s/\chi$. These plots confirm that the net force can be a propulsion (red region on the electronic version of this manuscript).

As expected from the results of \citetalias{2019MNRAS.483.4383V}, the higher the Mach number, the larger the luminosity required to get a net propulsion. The heating force needs indeed to overcome not only the cold force (roughly constant in the limit of a low Mach number) but also the adiabatic dynamical friction \citep{1999ApJ...513..252O}, which increases with the Mach number. We see indeed that the luminosity required to cancel the force (i.e. the watershed between a net drag --in blue in the electronic version-- and a net propulsion --in red in the electronic version, shown by a thick solid line on the plots) is $\sim 2L_c$ for ${\cal M}=0.4$ for the lowest masses, whereas is it only marginally larger than $L_c$ for ${\cal M}=0.1$.
From now on we shall call sub-critical masses those which obey $M<M_c$, and super-critical masses those with $M>M_c$. We find that for sub-critical masses the critical luminosity that separates drag from propulsion is roughly independent of the mass, in agreement with the results of \citetalias{2019MNRAS.483.4383V}, whereas for super-critical masses it becomes increasingly difficult to get a propulsion as the mass increases: the heating force has therefore, for these masses, a value smaller than that of the analytic expression of Eq.~\eqref{eq:9}, even more so as the mass increases. From now on we refer to this effect as the cutoff of the force for super-critical masses.

The extent of our numerical exploration is limited, on the left and upper sides of the plots, by the growth in computational cost as we explore smaller masses and larger luminosities. On the one hand, the necessity of resolving the Bondi radius $R_B$ imposes a quadratic growth on the number of timesteps as we reduce $R_B$ with a fixed $R_B/\Delta z$ ratio: for sub-critical masses, almost by definition, the timestep is limited by thermal diffusion rather than by advection or acoustic waves. The maximum allowed timestep therefore scales quadratically with the resolution in this regime, instead of linearly, making it very expensive to simulate low masses for a fixed number of cells per Bondi radius. On the other hand, for super-critical luminosities, the increased temperature in the Bondi sphere implies an increases thermal diffusivity, as per Eq.~\eqref{eq:17}, and doubling the luminosity roughly amounts to doubling the execution time of the simulation.

Finally, since the time to establish the thermal forces scales as ${\cal M}^{-2}$, as can be seen from Eqs.~\eqref{eq:15} and~\eqref{eq:16}, simulations at low Mach numbers are very computationally expensive. Although it would be desirable to tackle Mach numbers as low as possible in order to enable the cleanest possible comparison with analytic results (which are obtained in the limit of a vanishing Mach number), a trade off must be made between the value of the Mach number and the extent of our exploration of the parameter space.

\begin{figure}
\includegraphics[width=\columnwidth]{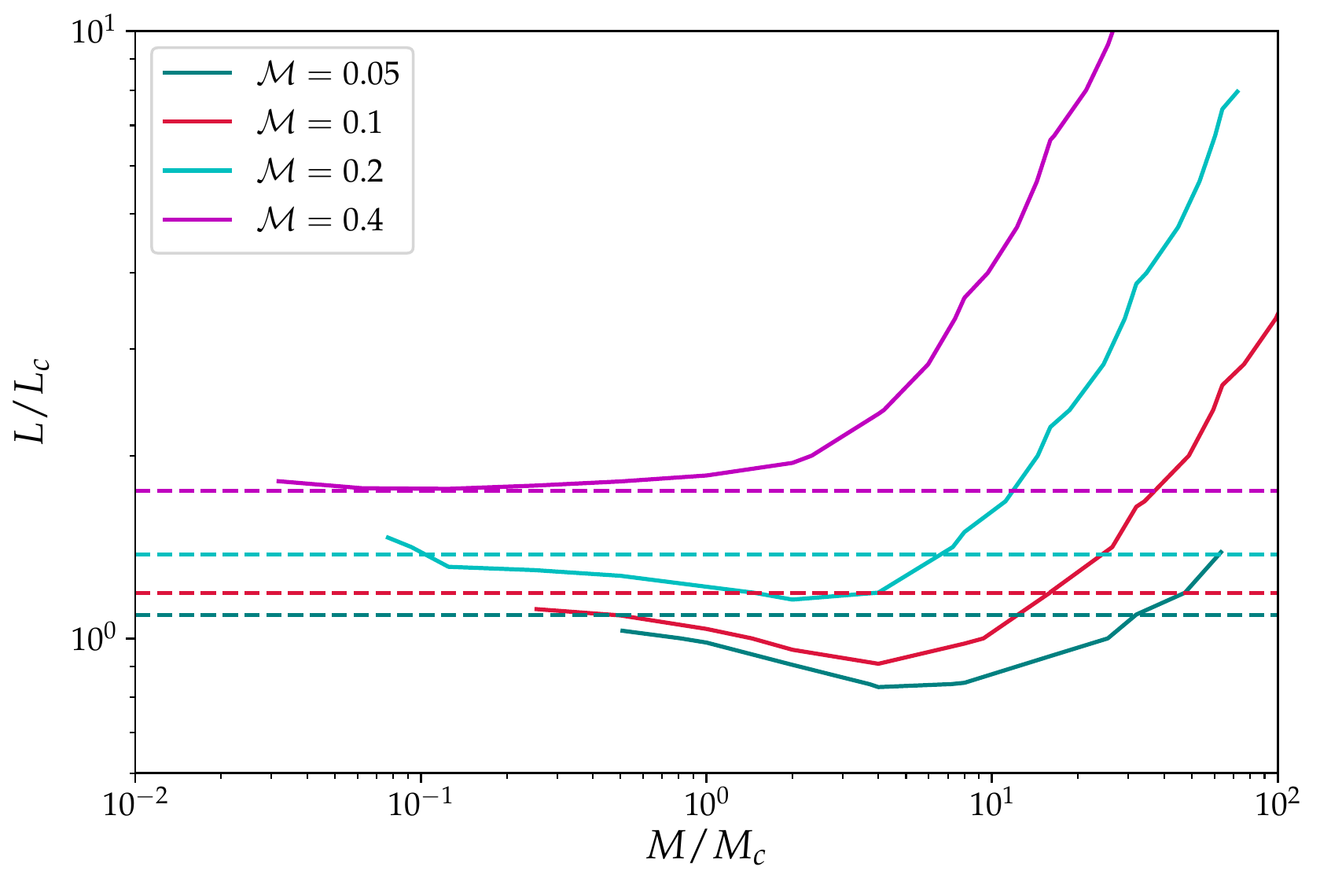}
 \caption{Luminosity required to cancel the net force, as a function of the mass of the perturber, for  different Mach numbers. In dashed lines we show the values obtained from Eq.~\eqref{eq:22}.}
\label{fig:2}
\end{figure}

In Fig.~\ref{fig:2} we plot the luminosity required to get a null net force as a function of the mass of the perturber, for different values of the Mach number. The purple and red curves, in the electronic version of this manuscript, correspond to the solid thick lines of, respectively, the left and right plots of Fig.~\ref{fig:1}. Assuming that the Taylor expansion of the thermal forces in ${\cal M}$ does not contain first order term in the vicinity of ${\cal M}=0$ (a reasonable assumption motivated by the fact that all curves of Fig.~14 of \citetalias{2019MNRAS.483.4383V} have very similar slopes at the origin, so that the first order term in ${\cal M}$ that dominates the expression of the total force is that of the adiabatic force), we can write to first order in ${\cal M}$ the total force using Eqs.~\eqref{eq:7} to~\eqref{eq:12} as
\begin{equation}
  \label{eq:21}
  F_\mathrm{total}=\left[\frac{\gamma-1}{2}\left(\frac{L}{L_c}-1\right)-\frac{\cal M}{3}\right]F_0.
\end{equation}
We therefore expect this net force to cancel out for:
\begin{equation}
  \label{eq:22}
  \frac{L}{L_c}=\frac{2{\cal M}}{3(\gamma-1)}+1=\frac 53{\cal M}+1.
\end{equation}
The horizontal dashed lines in Fig.~\ref{fig:2} show this critical value of the luminosity for the different Mach numbers. We note that for the three smaller values of the Mach number, the curves intersect their corresponding horizontal dashed line at $M/M_c\sim 30$ (for ${\cal M}=0.05$), $M/M_c\sim 15$ (for ${\cal M}=0.1$) and $M/M_c\sim 7$ (for ${\cal M}=0.2$). We also note that below these values, the curves display a slight drop (by a factor of $\sim 2/3$ at most), so that a luminosity smaller than that expected from Eq.~\eqref{eq:21} suffices to reverse the net force from a drag to a propulsion. We remark that the masses quoted above obey the approximate relationship:
\begin{equation}
  \label{eq:23}
  \frac{M}{M_c}\sim \frac{1.5}{\cal M}.
\end{equation}
Using Eq.~\eqref{eq:14} and \eqref{eq:15}, we can infer that the mass of a perturber for which the net force changes sign for a luminosity approximately given by Eq.~\eqref{eq:22} (specifying to the case $\gamma=1.4$) must obey:
\begin{equation}
  \label{eq:24}
  \frac{GM}{c_s^2}\lesssim 2\lambda.
\end{equation}
In other words Eq.~\eqref{eq:22} gives a correct approximate value of luminosity required to change the sign of the net force as long as the Bondi radius of the planet is smaller than twice the cutoff distance $\lambda$.

\subsection{Thermal efficiency}
\label{sec:thermal-efficiency}
\begin{figure*}
\includegraphics[width=\textwidth]{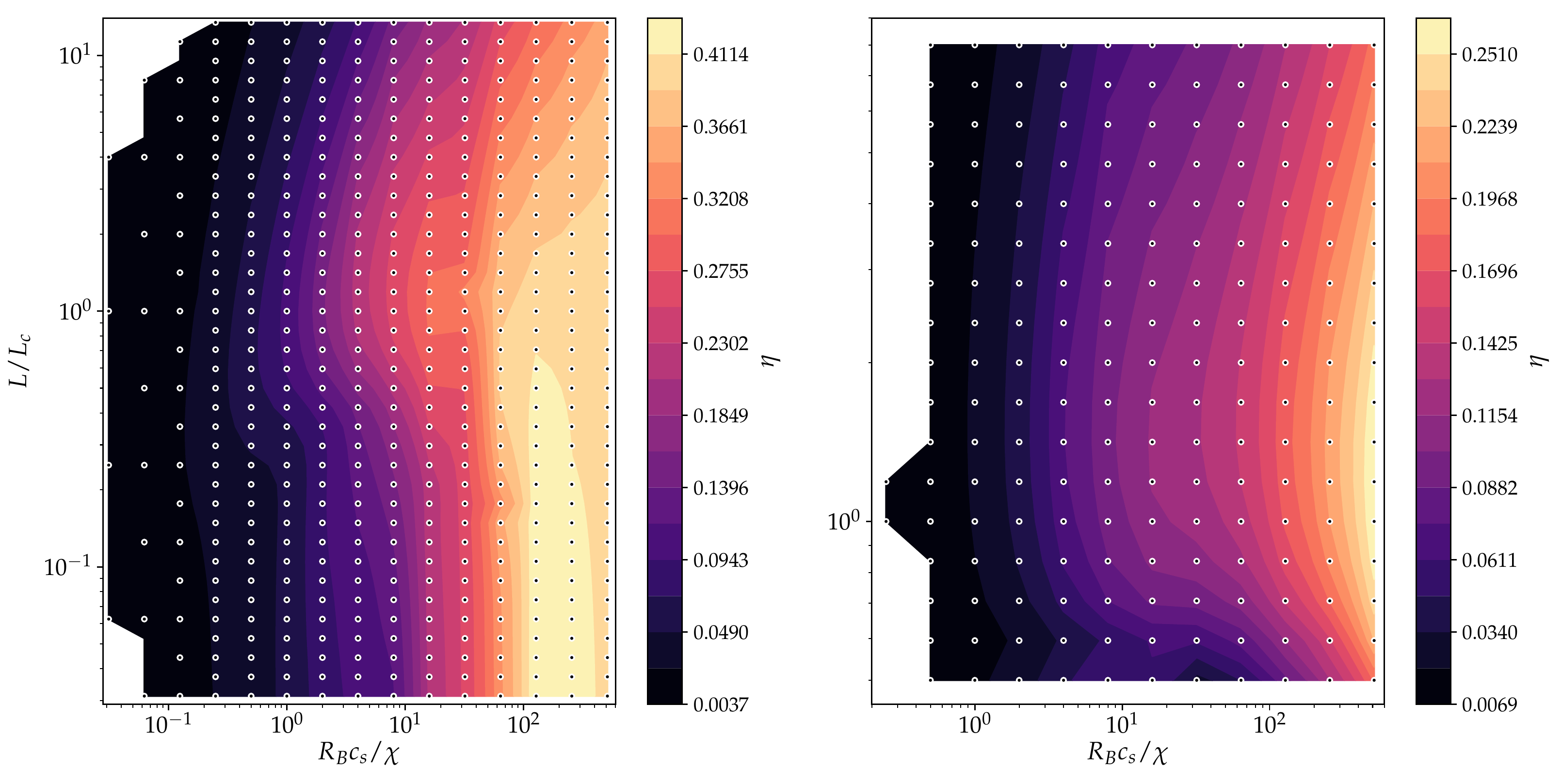}
\caption{Colour maps for the yield or thermal efficiency of the heating force. The left image shows results for ${\cal M} = 0.4$ while the right image shows the results for ${\cal M} = 0.1$. The $x-$axis is $R_Bc_s/\chi\equiv M/M_c$. On the low-mass side (i.e. on the left of the plots) we notice that the efficiency increases with the mass or the Bondi radius, in agreement with Eq.~\eqref{eq:25} or~\eqref{eq:26}, then saturates to values below one on the right side.}
\label{fig:3}
\end{figure*}

As derived by \citetalias{2017MNRAS.465.3175M}, in the limit of vanishing Mach number, the yield or thermal efficiency of the heating force $\eta=VF_\text{heating}/L$ reads:
\begin{equation}
  \label{eq:25}
  \eta=\frac{\gamma(\gamma-1)}{2}\left(\frac{GM}{c_s\chi}\right){\cal M},
\end{equation}
which can then be recast as
\begin{equation}
  \label{eq:26}
  \eta=\frac{(\gamma-1)}{2}\frac{R_B}{\lambda}.
\end{equation}
If the heating force were to always have the expression found by a linear analysis (Eq.~\ref{eq:5}), irrespectively of the mass of the perturber, Eq.~\eqref{eq:26} would yield a result higher than one for masses satisfying $R_B > 2\lambda/(\gamma-1)$. Therefore, there has to be a cutoff on the heating force for sufficiently large masses. Fig.~\ref{fig:3} shows two colour maps for the yield of the heating force corresponding to the simulations shown in Fig.~\ref{fig:1} (the heating force being defined as in previous work as the force difference between a simulation with a luminous planet and another simulation with same parameters, except that the planet is non-luminous). Here we can observe that indeed, the thermal efficiency is bound by $\sim 0.4$ for the case of a Mach number ${\cal M}=0.4$ and by $\sim 0.25$ for the case of a Mach number ${\cal M}=0.1$ (although in this last case the plot suggests that the efficiency could be higher for masses larger than those considered in our exploration of the parameter space).

\subsection{Cold force}
\label{sec:cold-force}
\begin{figure} \includegraphics[width=\columnwidth]{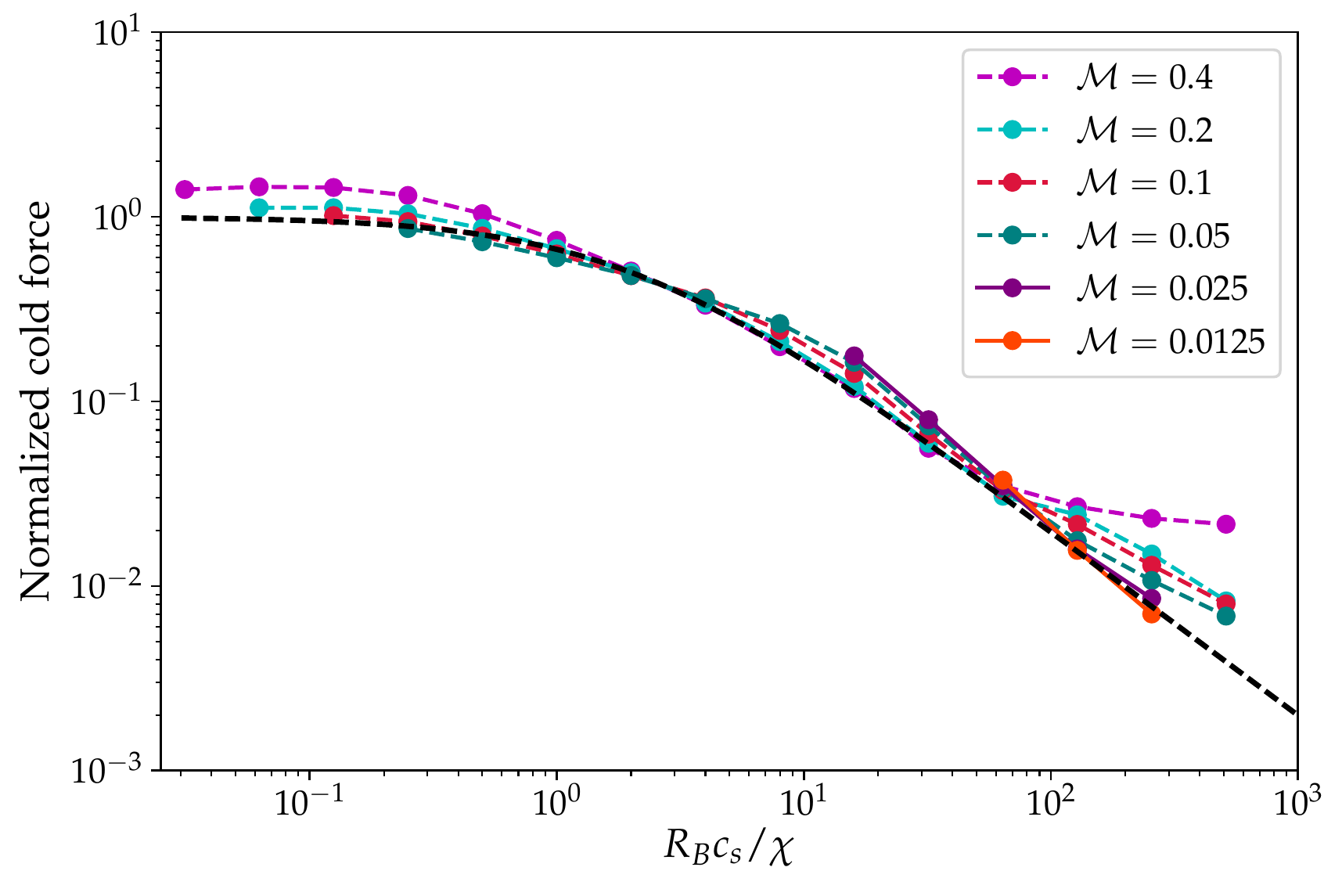} \caption{Cold force as a function of the mass for different Mach numbers. The forces are normalized by the asymptotic value at low Mach number obtained by a linear analysis (Eq.~\ref{eq:4} and \citetalias{2019MNRAS.483.4383V}) for the respective mass. We observe a cutoff of the force for super-critical masses ($R_Bc_s/\chi>1$ or equivalently $M>M_c$). The black dashed line corresponds to the cutoff formula for the cold force presented in section~\ref{sec:cut-formula}.}
\label{fig:4}
\end{figure}

In Fig.~\ref{fig:4} we show the cold force as a function of the mass for different values of the Mach number in the subsonic regime. Its value is obtained from the difference between the force measured in a simulation with thermal diffusion and 
the analytical expression of Eq.~\eqref{eq:10} for the same parameters in adiabatic scenario. Each value is normalized to the expression of Eq.~\eqref{eq:7}.
For sub-critical masses ($M<M_c$ or equivalently $R_Bc_s/\chi < 1$) we observe that the cold force is comparable to the value given by Eq.~\eqref{eq:4}, and that it can be marginally larger (by up to $\sim 30$~\% for the case of ${\cal M}=0.4$) than this value.
For super-critical masses ($R_Bc_s/\chi > 1$), we observe a cutoff on the thermal cold force: its value is significantly below that given by Eq.~\eqref{eq:4}, even more so as the mass (or Bondi radius) increases. In this regime, the cold force exhibits a dependence in the mass that is compatible with a scaling of $M^{-1}$ for the smallest Mach number, and a slightly more shallow relationship for the largest values of the Mach numbers used in our numerical exploration.

We mention that we include here simulations with Mach numbers smaller than those considered in the 2D exploration of the parameter space presented in sections~\ref{sec:net-force} and~\ref{sec:thermal-efficiency}. There are two reasons for that: first, only a relatively small number of such runs are needed in this exploration, and second, we do not perform simulations for the lowest mass values at low Mach numbers, as these would be too expensive.  Rather we focus on the asymptotic behaviour at large mass, which can be obtained at a moderate computational cost, in order to confirm the trends found for larger Mach numbers.

\subsection{Heating force}
\label{sec:heating-force}
\begin{figure}
\includegraphics[width=\columnwidth]{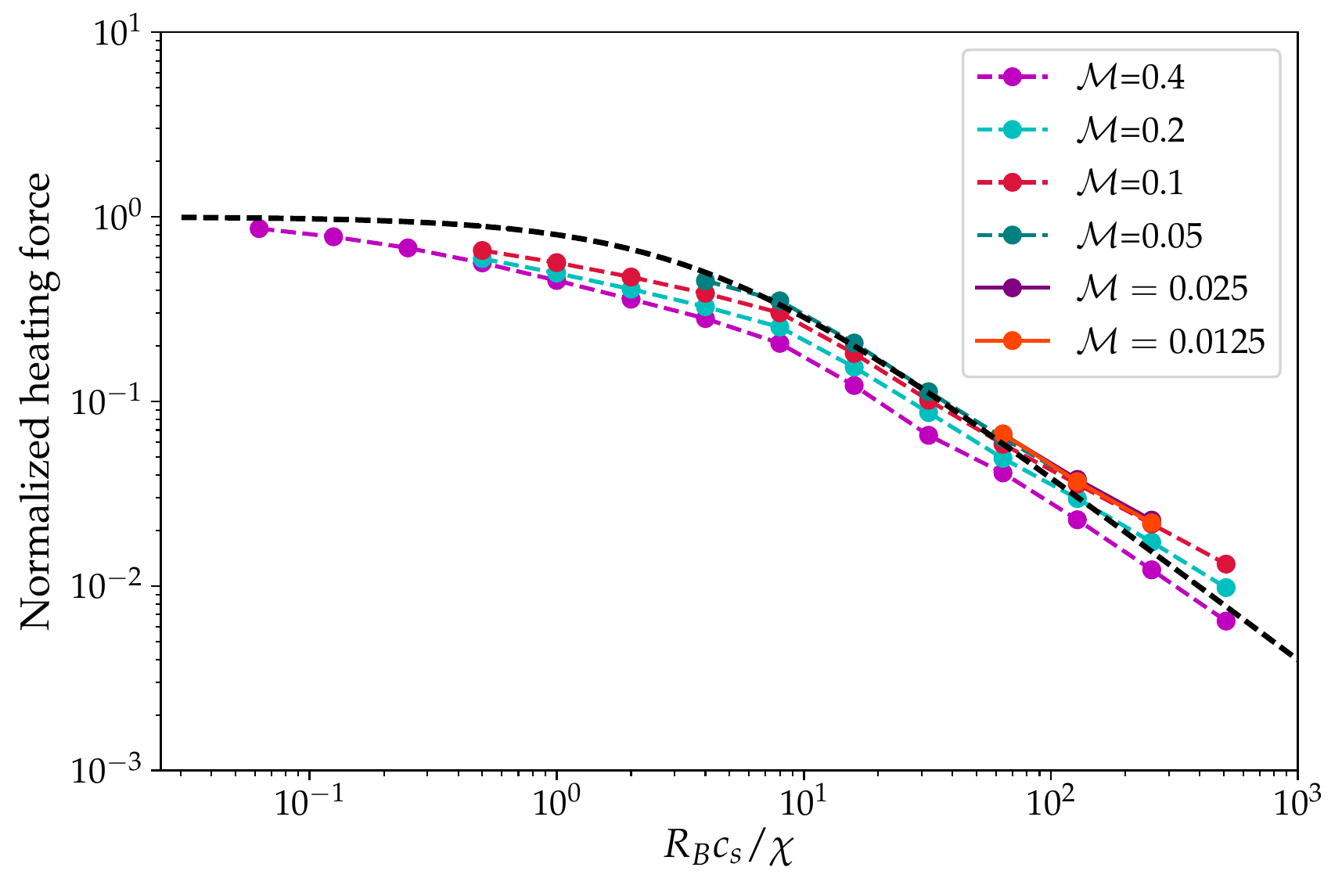}
\caption{Heating force for a luminosity of $L=8L_c$ as a function of the mass for different Mach numbers. The forces are normalized to the value given by Eq.~\eqref{eq:9} with $L/L_c=8$. As in Fig.~\ref{fig:4}, we observe a decay of the force value at larger mass. The black dashed line corresponds to the cutoff formula for the heating force presented in section~\ref{sec:cut-formula}.}
\label{fig:5}
\end{figure}

Now we proceed to analyse in a similar fashion the behaviour of the heating force as a function of the mass, again, for different values of the Mach number in the subsonic regime. In Fig.~\ref{fig:5} we plot the heating force for a super-critical luminosity of $L=8L_c$. For each point we evaluate the heating force as the difference between the force measured in a simulation with a hot perturber ($L=8L_c$) and one with a cold perturber ($L=0.0$), all other parameters of the simulations being the same. For all the curves presented here we observe a heating force lower than the asymptotic value of Eq.~\eqref{eq:5}, but only marginally so for lower mass (sub-critical) planets. For super-critical masses we observe that the heating force also presents a cutoff (in agreement with the expectations of section~\ref{sec:thermal-efficiency}). The decay of the force in this regime is compatible with a $\propto M^{-1}$ scaling for all Mach numbers. The curves of this plot show some offset relative to each other. In order to assess whether the $x-$axis variable $R_Bc_s/\chi\equiv M/M_c$ is a relevant choice, we have performed subsidiary runs at very low Mach number for super-critical masses. Those are found to overlap with the results for ${\cal M}=0.1$ and ${\cal M}=0.05$, suggesting that the scaling found for super-critical masses is independent indeed of the Mach number when the latter is vanishingly small.

\subsection{Bondi sphere}
\label{sec:bondi-sphere}
\begin{figure*}
\includegraphics[width=\textwidth]{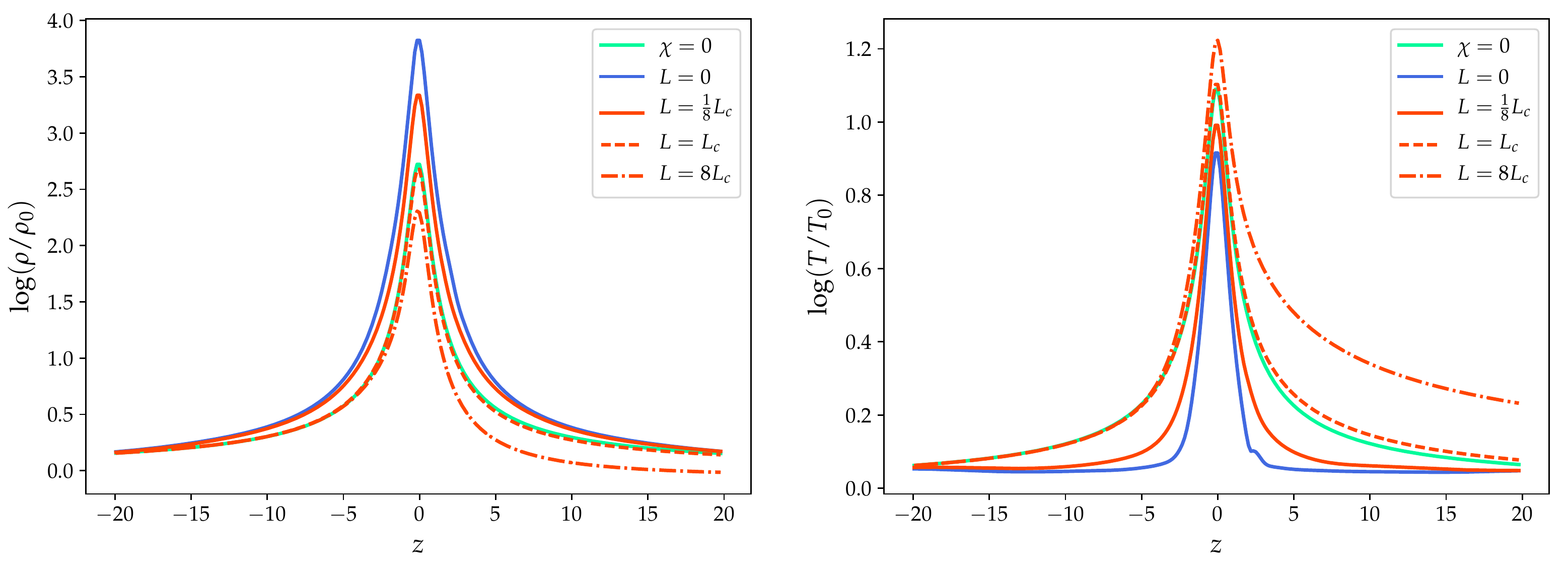}
\caption{Density cuts (left) and temperature cuts (right) along the $z-$axis at $t=200\tau$. In green we show the adiabatic case, in blue the cold case, and in orange the luminous case for three different values for the luminosity $L=1/8, 1$ and $8L_c$.}
\label{fig:6}
\end{figure*}

Here we focus on the dynamics inside the Bondi sphere of super-critical masses, and analyse the impact of the perturber's luminosity on the dynamics. Specifically, we present the results for a perturber of mass $M=32M_c$ with a Mach number of ${\cal M}=0.05$. In Fig.~\ref{fig:6} we show cuts of density and temperature along the $z$~axis for five different cases (the adiabatic case, the cold case and three cases with a luminous perturber, respectively with $L=L_c/8$, $L=L_c$ and $L=8L_c$). 

By comparing the adiabatic and cold case, we can observe the impact of thermal diffusion on both density and temperature: the former rises in the vicinity of the perturber, while the peak of the latter decreases around the perturber. As mentioned in \citetalias{2019MNRAS.483.4383V}, we observe that the perturber's luminosity tends to counteract this effect.  In particular, we recover a behaviour closely similar to that of the adiabatic case for the critical luminosity $L_c$. For the case of a super-critical luminosity, we observe an asymmetry between the front and rear regions of the perturber, both in temperature and density. The gas, in particular, is underdense behind the perturber (i.e. for $z>0$), in agreement with the fact that for the pair $(M,L)=(32M_c,8L_c)$ and a Mach number ${\cal M}=0.05$, the perturber experiences a net propulsion (see Fig.~\ref{fig:2}).

We now make use of the passive tracing field $\Psi=z_0$ to observe the whereabouts of the fluid elements during the simulation.  For $L=L_c$ the temperature and density fields nearly coincide with those of the adiabatic case, which corresponds to a uniform entropy, hence we expect the entropy gradient to vanish for $L\approx L_c$ in the vicinity of the perturber. For $L>L_c$, the gradient of temperature is super-adiabatic, and conversely the gradient of density is sub-adiabatic, so that there is a net gradient of entropy outwards from the perturber, hence anti-aligned with the gravity field $-\nabla\Phi$. The opposite holds for $L<L_c$, in which case the entropy gradient in the vicinity of the perturber is aligned with the gravity field.  If the perturber were at rest and thermal diffusivity would be negligible, the case $L=L_c$ would therefore correspond to the onset of convection: the gas surrounding the perturber would be stable for $L<L_c$, and unstable to convection for $L>L_c$. The present setup is different from this idealised situation, hence the flow properties are also different. However, as we shall see in the following, there is a clear change in the properties of the flow for a luminosity of the order of magnitude of $L_c$.

\begin{figure*}
\includegraphics[width=\textwidth]{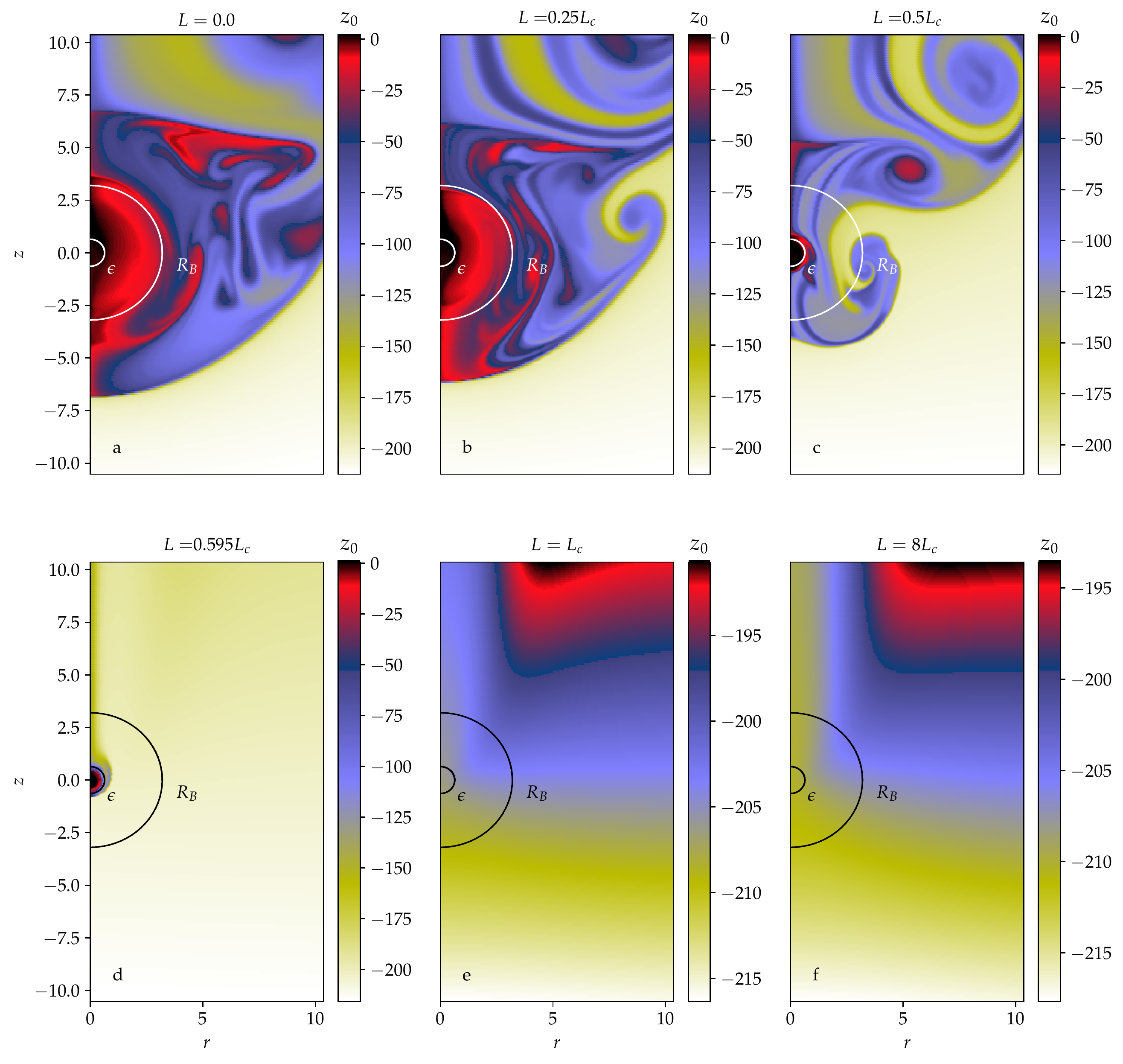}
\caption{Colour maps of the passive tracing field $\Psi=z_0$. The images correspond to a sequence of simulations with ${\cal M} = 0.1$ and $M=32M_c$ at $t=200\tau$, for increasing luminosities $L=0, 0.25, 0.5, 0.595, 1$ and $8L_c$. We draw the reader's attention to the very different colour scales (in the electronic version of this manuscript) for the plots a to d on the one hand, and for the plots e and f on the other hand. The former have a maximum value near zero, revealing the existence of material trapped since the beginning of the simulation, whereas the latter have values around $\sim -200$, which entails that all the material present on those maps was located much further upstream at the beginning of the simulation, precluding the existence of trapped material. The concentric half circles on these plots represent the smoothing length of the potential and Bondi's radius.}
\label{fig:7}
\end{figure*}

In Fig.~\ref{fig:7} we present colour maps of the tracing field for ${\cal M} = 0.1$ and $M=32M_c$  at $t=200\tau$. Over this timescale, the wind sweeps a distance $d$ given by:
\begin{equation}
  \label{eq:27}
  \frac{d}{R_B}=200\left(\frac{M}{M_c}\right)^{-1}\frac{1}{{\cal M}\gamma^2}\approx 64,
\end{equation}
thus fluid elements on  these snapshots were initially at distances almost two orders of magnitude larger than the Bondi radius from the perturber.
In Fig.~7a, we can observe entrapped material in a radius close to $R_B$. It is the material that appears in red or black colour in the electronic version of this manuscript, as it is material that was initially at $z$-values comparable to that of the perturber. In the images $b$ and $c$ of Fig.~\ref{fig:7}, we present respectively the cases for a perturber with luminosity $L_c/4    $ and $L_c/2$. We can see here that the overall behaviour deviates little from the one observed in the cold case. For a luminosity of $L=0.595L_c$ (shown in fig~\ref{fig:7}d), we now observe a qualitatively different behaviour from the cold case, with a large reduction in the size of the region with entrapped material. For the critical luminosity (Fig.~\ref{fig:7}e) we see that the material no longer remains trapped in the perturber's vicinity. In fact, the material that passes by the perturber is accelerated, as can be seen by the trail behind the perturber ($z>0, r<1-2$). A similar behaviour is also observed in Fig.~\ref{fig:7}f where we present the case for $L=8L_c$. We finally mention that the critical luminosity $L_c$ used here has been estimated using the unperturbed value of $\chi$. Owing to the strong perturbations of density and temperature within the Bondi radius, the effective value of $\chi$ in that region is substantially smaller than the unperturbed one, by virtue of Eq.~\eqref{eq:17}.

\section{A simple cutoff model}
\label{sec:simple-cut-model}
\subsection{Considerations from linear theory}
\label{sec:gener-cons}
We consider here the heating force exerted by the gas located outside of the Bondi radius onto the perturber. For luminosities small enough that the perturbation of density associated to the heat release is small outside of the Bondi radius, this corresponds to the regions of the flow where a linearisation of the governing equations is sensible. Using Eqs.~(28) and~(32) of \citetalias{2017MNRAS.465.3175M}, we can express this force as:
\begin{eqnarray}
  F_c&=&F_\mathrm{heating}\left[h(+\infty)-h\left(\frac{R_B}{2\lambda}\right)\right]\nonumber\\
 &=&F_\mathrm{heating}\frac{2R_B'+\exp(-2R_B')-1}{2R_B'^2},   \label{eq:28}
\end{eqnarray}
where $R_B'\equiv R_B/2\lambda$ and $F_\mathrm{heating}$ is given by Eq.~\eqref{eq:9}. The argument of the function $h$ that corresponds to the radius of exclusion must be divided by $2\lambda$ \citepalias{2017MNRAS.465.3175M}, where $\lambda$ is given by Eq.~\eqref{eq:15}, in order to be dimensionless. Such an expression (or an expression closely related in which $R_B'$ is substituted by $bR_B'$, where $b>1$) applies for instance to the case of an energy release directly outside of the Bondi sphere (if, for instance, the mean free path of the photons emitted by the perturber is larger than the Bondi radius). In such case, linear theory can be used to evaluate the force regardless of the perturber's mass (as long as the plume size is much larger than the mean free path of the photons). The yield of the thermal force, in particular, is:
\begin{equation}
  \label{eq:29}
  \eta=(\gamma-1) \frac{2R_B'+\exp(-2R_B')-1}{2R_B'},
\end{equation}
and is bound by $\gamma-1$. Eq.~\eqref{eq:28} entails:
\begin{equation}
  \label{eq:30}
  F_c\approx \begin{cases}
    F_\mathrm{heating}\mbox{~~for ${\cal M}M/M_c\ll 1$}\\
    F_\mathrm{heating}\frac{2}{\gamma}\frac{M_c}{M}{\cal M}^{-1}\mbox{~~for ${\cal M}M/M_c\gg 1$}\\
  \end{cases}
\end{equation}
In the simulations performed in this work, the energy release occurs well inside the Bondi radius (see section~\ref{sec:heat-release-1}), so that the above considerations do not apply directly. However, the forces obtained in our numerical experiments, which show a cutoff around $M_\text{cutoff}\sim M_c$, are compatible with those provided by the  analytic expressions obtained from a linear analysis, in which the contribution from central regions is omitted. The extent of the central regions to be omitted differs however from the Bondi radius.
As we saw above, omitting a region of the extent of the Bondi radius results in a cutoff that occurs around $M_\text{cutoff}\sim M_c{\cal M}^{-1}$.
The numerical experiments presented in sections~\ref{sec:cold-force} and~\ref{sec:heating-force}, which focus respectively on the cutoff of the cold force and that of the heating force for $L=8L_c$, are well reproduced by excising a central region of size $\propto R_B/{\cal M}$ instead of $R_B$. We shall determine the best size for the heating and cold forces in section~\ref{sec:cut-formula}.

While the cutoff of the heating force was expected from first principles, that of the cold force was not. It can be understood as follows. As we saw in section~\ref{sec:bondi-sphere}, material circulates within the Bondi sphere of a non-luminous perturber of super-critical mass. It does so at velocities $\sim c_s$. The timescale required to build the enthalpy peak around the perturber is therefore $R_B/c_s$, while the time it takes to smooth it by thermal diffusion is $R_B^2/\chi$. When this timescale is longer than the former (i.e., when $M>M_c$), a significant enthalpy peak still subsists around the perturber (in agreement with the cuts presented in Fig.~\ref{fig:6}), and the cold force is smaller than its nominal value of Eq.~\eqref{eq:4}.

\subsection{A cutoff formula}
\label{sec:cut-formula}
\begin{figure}
  \centering
  \includegraphics[width=\columnwidth]{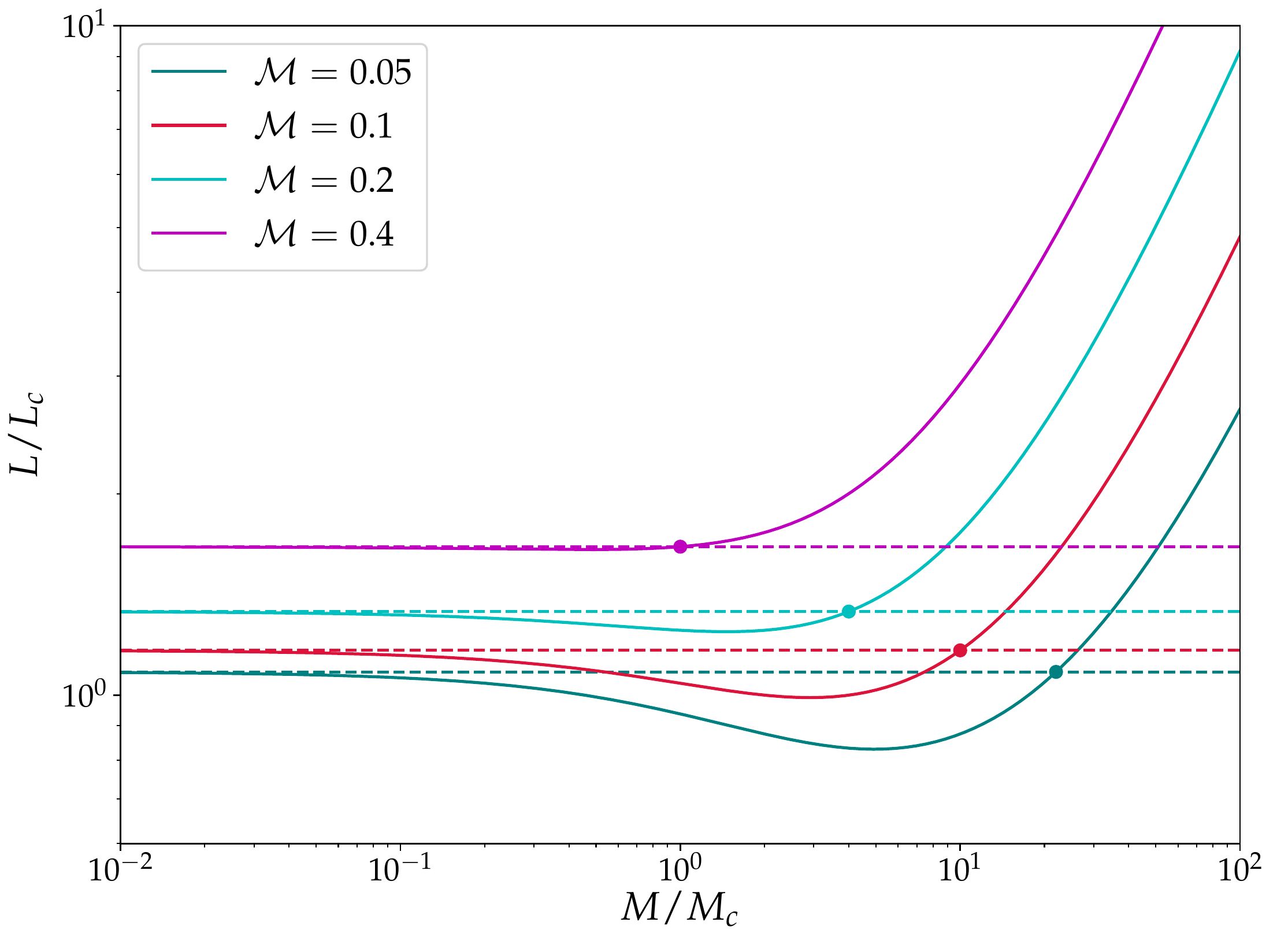}
  \caption{\label{fig:8}Luminosity required to cancel the net force as a function of the perturber's mass, for different Mach numbers, as given by Eq.~\eqref{eq:35}. This figure should be compared to Fig.~\ref{fig:2}.}
\end{figure}

We introduce the ``cutoff function'' $c$ defined by:
\begin{equation}
  \label{eq:32}
  c:x\mapsto \frac{2x+\exp(-2x)-1}{2x^2},  
\end{equation}
so that the heating force of Eq.~\eqref{eq:28} can be simply written as $F_\mathrm{heating}c(R_B')$. This function is nearly equal to the much simpler function:
\begin{equation}
  \label{eq:33}
  C:x\mapsto\frac{1}{1+x}.
\end{equation}
The functions $C$ and $c$ both tend to $1$ for $x\rightarrow 0$ and are both equivalent to $x^{-1}$ for $x\rightarrow+\infty$. They differ by at most $14$~\% in the vicinity of $x=1$. Given the dispersion of the curves observed on Figs.~\ref{fig:4} and~\ref{fig:5}, there is no need to handle the more complex function $c(x)$ and we use from now on the function $C(x)$ to describe the cutoffs. As mentioned in section~\ref{sec:gener-cons}, the size of the region to be omitted should scale with $R_B/{\cal M}$, hence the argument of the function $C$ should scale with $R_B/(2\lambda{\cal M})\,\propto\, M/M_c$. We find that different arguments should be used for the cold and heating forces. Using the data of Fig.~\ref{fig:4}, we find that the cutoff that best fits the cold force is $C(M/2M_c)$, whereas from the data for Fig.~\ref{fig:5} we find that the cutoff that best fits the heating force is $C(M/4M_c)$. The factors $2$ and $4$ in the denominator of the argument are given with a one digit accuracy owing to the significant dispersion of the curves of Figs.~\ref{fig:4} and~\ref{fig:5}. The cold force is therefore cut off at slightly smaller mass than the thermal force, which explains the drop in the zero-force luminosity found in Fig.~\ref{fig:2}, as we shall see in further detail below.
We can now provide the following expression for the net force:
\begin{equation}
  \label{eq:34}
  F_\mathrm{total}=F_\mathrm{adi}+F_\mathrm{thermal}^\mathrm{cold}\frac{2M_c}{M+2M_c}+F_\mathrm{heating}\frac{4M_c}{M+4M_c},
\end{equation}
which yields same result as Eq.~\eqref{eq:12} in the limit $M\ll M_c$. This expression gives results compatible with those presented in section~\ref{sec:net-force} about the luminosity required to cancel the net force. Using a first order expansion of the force in the Mach number, we find that the luminosity corresponding to a zero net force reads:
\begin{equation}
  \label{eq:35}
\frac{L}{L_c}=\left[\frac{2{\cal M}}{3(\gamma-1)}+\frac{1}{1+M/2M_c}\right]\left(1+\frac{M}{4M_c}\right).
\end{equation}
This relationship is plotted in Fig.~\ref{fig:8}, where we recover several features of Fig.~\ref{fig:2}, which has been obtained directly from numerical simulations. Namely, we recover (i) the existence of a minimum below the asymptotic value at low mass, the drop being more important for low Mach numbers, and virtually nonexistent for ${\cal M}=0.4$  and (ii) the fact that the curves intersect their asymptote at larger values of $M/M_c$ for lower Mach number (as represented by the dots on this figure). More precisely, Eq.~\eqref{eq:35} implies that the curves intersect their asymptote for:
\begin{equation}
  \label{eq:36}
  \frac{M}{M_c}=\frac{3(\gamma-1)}{\cal M}-2=\frac{1.2}{\cal M}-2.
\end{equation}
This relationship is broadly consistent with that found in Eq.~\eqref{eq:23} in the limit of a low Mach number.

\section{Application to low mass embryos in a protoplanetary disc}
We now turn to the evaluation of the luminosity required to get a net propulsion on low-mass embryos embedded in protoplanetary discs. We stress that the present analysis cannot be applied to giant planets. Those are surrounded by circumplanetary discs which have a size comparable to the pressure length scale of the disc. Only for eccentricities comparable to or larger than the aspect ratio of the disc can these circumplanetary discs have a much smaller size. This requires, however, that the planet is supersonic with respect to the disc, which falls beyond the range of validity o the present work.

\begin{figure*}
  \centering
  \includegraphics[width=\textwidth]{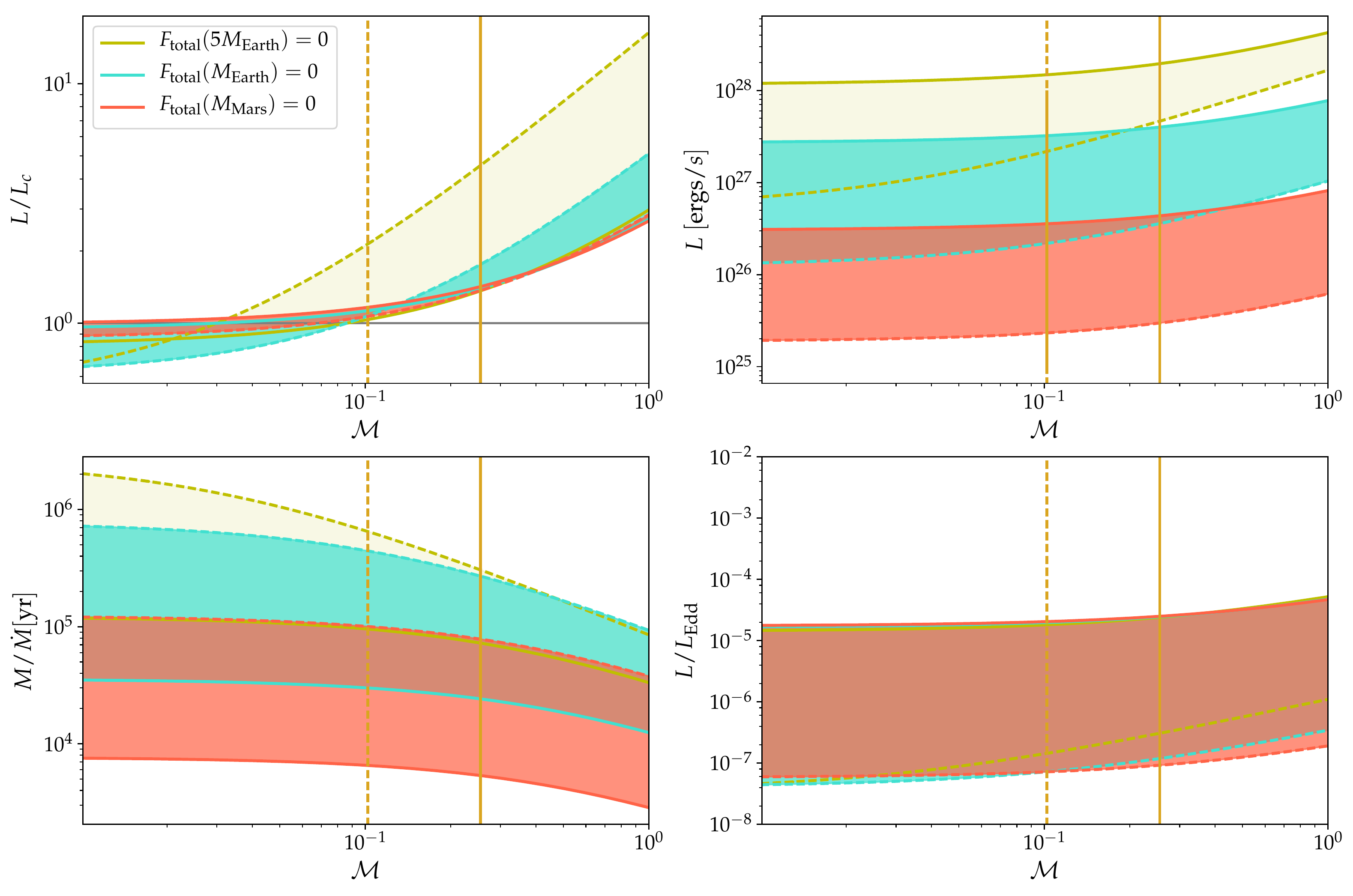}
  \caption{\label{fig:9}Luminosity required to cancel the net force as a function of the Mach number for two different planet masses and at different stages of the disc's evolution, as given by Table.~\ref{tab:1} for the disc model of \citet{2015A&A...582A.112B}. See text for details.}
\end{figure*}

We return to the values presented in Table.~\ref{tab:1} for the disc model of \citet{2015A&A...582A.112B} at two different stages of its evolution ($t=300\;\text{kyr}$ and $t=1\;\text{Myr}$). In Fig.~\ref{fig:9} we present a number of quantities derived from the luminosity required to have a null net force, as a function of the Mach number, for three different planet masses: one Mars mass, one Earth mass and five Earth masses. The solid lines correspond to the case  $t=300\;\text{kyr}$ while the dashed lines depict the late case at $t=1\;\text{Myr}$. The top-left picture shows the luminosity to critical luminosity ratio. As expected, it tends towards unity at low Mach number, except for the most massive embryo, for which a subcritical luminosity may suffice to bring the net force to a positive value, as we saw in Fig.~\ref{fig:8}. The top right figure shows the zero-force luminosity in units of erg/s,  the bottom-left plot shows the corresponding mass doubling time $M/\dot M$, evaluated assuming that the luminosity has the value \citep{2015Natur.520...63B}:
\begin{equation}
  \label{eq:39}
  L=\frac{GM\dot M}{R},
\end{equation}
where $R$ is the physical radius of  the planet, given by:
\begin{equation}
  \label{eq:40}
  R=\left(\frac{3M}{4\pi\rho}\right)^{1/3},
\end{equation}
where we have assumed $\rho=3$~g.cm$^{-3}$. On this plot we see that for a Mars mass a mass doubling time of $\sim10^4$~--~$10^5$ years is required to cancel out the adiabatic and cold drag forces, while for a one Earth mass (five Earth mass) core a mass doubling time five times longer (ten times longer) suffices.  Finally, the bottom right plot shows the ratio of the zero-force luminosity to the Eddington luminosity: \begin{equation}
  \label{eq:37}
{L_\text{Edd}}= \frac{4\pi G M c}{\kappa},
\end{equation}
where $c$ is the speed of light. This ratio is always much smaller than one in the context of protoplanetary discs.
Indeed, using Eqs.~\eqref{eq:6} and~\eqref{eq:37}, the ratio of the critical to Eddington luminosities can be expressed as:
\begin{equation}
  \label{eq:38}
\frac{L_c}{L_\text{Edd}}= \frac{\rho_0 \kappa \chi }{\gamma c} \sim \frac{\chi}{c\bar{\ell}},
\end{equation}
where we neglect dimensionless factors of order of unity (such as $\gamma$) in this comparison of orders of magnitude. The thermal diffusivity in the material of high optical depth surrounding the embryo can be expressed as \citep[see e.g.][]{pbk11}:
\begin{equation}
  \label{eq:41}
  \chi\sim\frac{\sigma T^4}{\kappa \rho P}\sim\frac{\sigma T^4\bar{\ell}}{P},
\end{equation}
where $P$ is the gas pressure. Using the radiation constant $a=4\sigma/c$, Eq.~\eqref{eq:41} can be recast as:
\begin{equation}
  \label{eq:42}
  \chi\sim c\bar\ell\cdot\frac{aT^4}{P}\sim c\bar\ell\cdot\frac{P_\mathrm{rad}}{P}.
\end{equation}
Substituting this result into Eq.~\eqref{eq:38}, we are left with:
\begin{equation}
  \label{eq:43}
  \frac{L_c}{L_\mathrm{Edd}}\sim\frac{P_\mathrm{rad}}{P}.
\end{equation}
In the cold, dense environments of protoplanetary discs, the radiation pressure is usually orders of magnitude smaller than the gas pressure, hence the critical and zero-force luminosities are also orders of magnitude smaller than the Eddington luminosity, in agreement with Fig.~\ref{fig:9}. Such a statement may not hold in a different context, as for instance that of the dynamical friction on a massive black hole.

In the plots of Fig.~\ref{fig:9} the vertical lines show the minimum Mach number above which the thermal force acting on the planet can be adequately described with a dynamical friction calculation, respectively at $t=300$~kyr (solid line) and at $t=1$~Myr (dashed line). This minimal Mach number is found by the requirement that the response time of the hot trail $\chi/v^2$ is smaller than the shear timescale $\Omega^{-1}$, where $\Omega$ is the orbital frequency \citep{2017arXiv170401931E,2019MNRAS.485.5035F}. This requirement can be cast as:
\begin{equation}
  \label{eq:44}
  {\cal M} > \frac rH\sqrt{\frac{\chi}{r^2\Omega}},
\end{equation}
where $r$ is the distance to the star.

\section{Conclusions}
\label{sec:disc-concl}
We have performed high resolution simulations in the subsonic regime in two and three dimensions that resolve the Bondi radius of a perturber travelling at constant speed in a gas with initially uniform density and temperature. We have found that for perturbers with sufficiently small mass, the net force is in agreement with that obtained from linear theory \citepalias{2017MNRAS.465.3175M,2019MNRAS.483.4383V}. In particular, we find a net propulsion for perturbers with luminosities $L\gtrsim L_c[1+(5/3){\cal M}]$, where $L_c$ is given by Eq.~\eqref{eq:6} and ${\cal M}$ is the Mach number. At larger masses, the thermal forces are smaller than those given by linear theory (this is the case both for the heating force or radiative feedback onto the perturber, and for the cold force, which is the additional drag experienced by a non-luminous perturber in a gas with thermal diffusivity, with respect to the adiabatic drag). In this high-mass regime, the ratio of the thermal forces measured to those inferred from linear theory decays as $M^{-1}$ ($M$ being the perturber's mass), so that even at larger masses, it is possible to obtain a net propulsion, provided that the perturber has a high enough luminosity. The luminosity required, however, increases sharply with the perturber's mass. The transition between the low-mass and high-mass regimes (where the thermal force is $50$~\% of the linear estimate), occurs for a mass that is independent of the Mach number, and which is $\sim 2M_c$ for the cold force and $\sim 4M_c$ for the heating force, $M_c$ being given by:
\begin{equation}
  \label{eq:45}
  M_c=\frac{\chi c_s}{G},
\end{equation}
where $\chi$ is the thermal diffusivity, $c_s$ the adiabatic sound speed and $G$ the gravitational constant. The fact that the cold force is cut off at slightly lower mass than the heating force allows perturbers in the range $M_c$---$10\;M_c$ to experience a net propulsion even at subcritical luminosities.
The yield or thermal efficiency of the heating force increases with the perturber's mass, and saturates at values $\lesssim 1$ for bodies of super-critical mass.
We give in section~\ref{sec:simple-cut-model} simple, approximate laws for the cutoff of the cold force and heating force. We provide in Eq.~\eqref{eq:34} a formula for the net force valid both for sub- and super-critical masses.

We find that the flow around the perturber can be complex and highly time-variable, and that it has different topologies for low (sub-critical) and high (super-critical) luminosities. Despite this complexity, the net force is nearly constant in time and varies smoothly as the perturber's luminosity increases.

The astrophysical applications of the present study include moderately eccentric or inclined planetary embryos embedded in gaseous protoplanetary discs \citep{2017arXiv170401931E,2019MNRAS.485.5035F} or the dynamical friction on a massive black hole after a galactic merger event \citep{2017ApJ...838..103P,2019ApJ...883..209P}.

The analysis presented here assumes that heat transfer between gas parcels obeys a diffusion equation such as that of Eq.~\eqref{eq:3}. When heat diffusion is effected by radiative transfer, this assumption is reasonable as long as the mean free path of the photons is much smaller than the plume size and the Bondi radius of the perturber. When these assumptions break, more realistic calculations including radiative transfer should be undertaken. Also, the smoothing length of the potential adopted here is adequate for embryos typically Mars-sized. Describing larger mass embryos with a smoothing length of the potential comparable to the physical radius of the embryo would require numerical simulations with an even higher resolution than those presented here.

\section*{Acknowledgements}
D.V. acknowledges support from CONACyT's grant 412604. The authors acknowledge support from the Marcos Moshinsky Foundation through the Marcos Moshinsky Fellowship.




\bibliographystyle{mnras}
\bibliography{biblio} 

\begin{thebibliography}{}
\makeatletter
\relax
\def\mn@urlcharsother{\let\do\@makeother \do\$\do\&\do\#\do\^\do\_\do\%\do\~}
\def\mn@doi{\begingroup\mn@urlcharsother \@ifnextchar [ {\mn@doi@}
  {\mn@doi@[]}}
\def\mn@doi@[#1]#2{\def\@tempa{#1}\ifx\@tempa\@empty \href
  {http://dx.doi.org/#2} {doi:#2}\else \href {http://dx.doi.org/#2} {#1}\fi
  \endgroup}
\def\mn@eprint#1#2{\mn@eprint@#1:#2::\@nil}
\def\mn@eprint@arXiv#1{\href {http://arxiv.org/abs/#1} {{\tt arXiv:#1}}}
\def\mn@eprint@dblp#1{\href {http://dblp.uni-trier.de/rec/bibtex/#1.xml}
  {dblp:#1}}
\def\mn@eprint@#1:#2:#3:#4\@nil{\def\@tempa {#1}\def\@tempb {#2}\def\@tempc
  {#3}\ifx \@tempc \@empty \let \@tempc \@tempb \let \@tempb \@tempa \fi \ifx
  \@tempb \@empty \def\@tempb {arXiv}\fi \@ifundefined
  {mn@eprint@\@tempb}{\@tempb:\@tempc}{\expandafter \expandafter \csname
  mn@eprint@\@tempb\endcsname \expandafter{\@tempc}}}

\bibitem[\protect\citeauthoryear{{Ben\'{\i}tez-Llambay}, {Masset},
  {Koenigsberger}  \& {Szul\'agyi}}{{Ben\'{\i}tez-Llambay}
  et~al.}{2015}]{2015Natur.520...63B}
{Ben\'{\i}tez-Llambay} P.,  {Masset} F.,  {Koenigsberger} G.,   {Szul\'agyi}
  J.,  2015, \mn@doi [\nat] {10.1038/nature14277}, \href
  {http://adsabs.harvard.edu/abs/2015Natur.520...63B} {520, 63}

\bibitem[\protect\citeauthoryear{{Bitsch}, {Johansen}, {Lambrechts}  \&
  {Morbidelli}}{{Bitsch} et~al.}{2015a}]{2015A&A...575A..28B}
{Bitsch} B.,  {Johansen} A.,  {Lambrechts} M.,   {Morbidelli} A.,  2015a,
  \mn@doi [\aap] {10.1051/0004-6361/201424964}, \href
  {http://adsabs.harvard.edu/abs/2015A.26A...575A..28B} {575, A28}

\bibitem[\protect\citeauthoryear{{Bitsch}, {Lambrechts}  \&
  {Johansen}}{{Bitsch} et~al.}{2015b}]{2015A&A...582A.112B}
{Bitsch} B.,  {Lambrechts} M.,   {Johansen} A.,  2015b, \mn@doi [\aap]
  {10.1051/0004-6361/201526463}, \href
  {http://adsabs.harvard.edu/abs/2015A.26A...582A.112B} {582, A112}

\bibitem[\protect\citeauthoryear{{Cant{\'o}}, {Raga}, {Esquivel}  \&
  {S{\'a}nchez-Salcedo}}{{Cant{\'o}} et~al.}{2011}]{2011MNRAS.418.1238C}
{Cant{\'o}} J.,  {Raga} A.~C.,  {Esquivel} A.,   {S{\'a}nchez-Salcedo} F.~J.,
  2011, \mn@doi [\mnras] {10.1111/j.1365-2966.2011.19574.x}, \href
  {http://adsabs.harvard.edu/abs/2011MNRAS.418.1238C} {418, 1238}

\bibitem[\protect\citeauthoryear{{Eklund} \& {Masset}}{{Eklund} \&
  {Masset}}{2017}]{2017arXiv170401931E}
{Eklund} H.,  {Masset} F.~S.,  2017, \mn@doi [\mnras] {10.1093/mnras/stx856},
  \href {http://adsabs.harvard.edu/abs/2017MNRAS.469..206E} {469, 206}

\bibitem[\protect\citeauthoryear{{Fromenteau} \& {Masset}}{{Fromenteau} \&
  {Masset}}{2019}]{2019MNRAS.485.5035F}
{Fromenteau} S.,  {Masset} F.~S.,  2019, \mn@doi [\mnras]
  {10.1093/mnras/stz718}, \href
  {https://ui.adsabs.harvard.edu/abs/2019MNRAS.485.5035F} {485, 5035}

\bibitem[\protect\citeauthoryear{Kim \& Kim}{Kim \& Kim}{2009}]{Kim2009}
Kim H.,  Kim W.-T.,  2009, \mn@doi [The Astrophysical Journal]
  {10.1088/0004-637x/703/2/1278}, 703, 1278

\bibitem[\protect\citeauthoryear{{Kley} \& {Lin}}{{Kley} \&
  {Lin}}{1996}]{1996ApJ...461..933K}
{Kley} W.,  {Lin} D.~N.~C.,  1996, \mn@doi [\apj] {10.1086/177115}, \href
  {http://adsabs.harvard.edu/abs/1996ApJ...461..933K} {461, 933}

\bibitem[\protect\citeauthoryear{{Masset}}{{Masset}}{2017}]{2017MNRAS.472.4204M}
{Masset} F.~S.,  2017, \mn@doi [\mnras] {10.1093/mnras/stx2271}, \href
  {http://adsabs.harvard.edu/abs/2017MNRAS.472.4204M} {472, 4204}

\bibitem[\protect\citeauthoryear{{Masset} \& {Velasco Romero}}{{Masset} \&
  {Velasco Romero}}{2017}]{2017MNRAS.465.3175M}
{Masset} F.~S.,  {Velasco Romero} D.~A.,  2017, \mn@doi [\mnras]
  {10.1093/mnras/stw3008}, \href
  {http://adsabs.harvard.edu/abs/2017MNRAS.465.3175M} {465, 3175}

\bibitem[\protect\citeauthoryear{{Ostriker}}{{Ostriker}}{1999}]{1999ApJ...513..252O}
{Ostriker} E.~C.,  1999, \mn@doi [\apj] {10.1086/306858}, \href
  {http://adsabs.harvard.edu/abs/1999ApJ...513..252O} {513, 252}

\bibitem[\protect\citeauthoryear{{Paardekooper}, {Baruteau}  \&
  {Kley}}{{Paardekooper} et~al.}{2011}]{pbk11}
{Paardekooper} S.,  {Baruteau} C.,   {Kley} W.,  2011, \mn@doi [\mnras]
  {10.1111/j.1365-2966.2010.17442.x}, \href
  {http://adsabs.harvard.edu/abs/2011MNRAS.410..293P} {410, 293}

\bibitem[\protect\citeauthoryear{{Park} \& {Bogdanovi{\'c}}}{{Park} \&
  {Bogdanovi{\'c}}}{2017}]{2017ApJ...838..103P}
{Park} K.,  {Bogdanovi{\'c}} T.,  2017, \mn@doi [\apj]
  {10.3847/1538-4357/aa65ce}, \href
  {http://adsabs.harvard.edu/abs/2017ApJ...838..103P} {838, 103}

\bibitem[\protect\citeauthoryear{{Park} \& {Bogdanovi{\'c}}}{{Park} \&
  {Bogdanovi{\'c}}}{2019}]{2019ApJ...883..209P}
{Park} K.,  {Bogdanovi{\'c}} T.,  2019, \mn@doi [\apj]
  {10.3847/1538-4357/ab3f30}, \href
  {https://ui.adsabs.harvard.edu/abs/2019ApJ...883..209P} {883, 209}

\bibitem[\protect\citeauthoryear{{S{\'a}nchez-Salcedo} \&
  {Brandenburg}}{{S{\'a}nchez-Salcedo} \&
  {Brandenburg}}{1999}]{1999ApJ...522L..35S}
{S{\'a}nchez-Salcedo} F.~J.,  {Brandenburg} A.,  1999, \mn@doi [\apjl]
  {10.1086/312215}, \href {http://adsabs.harvard.edu/abs/1999ApJ...522L..35S}
  {522, L35}

\bibitem[\protect\citeauthoryear{{Schwarzschild}}{{Schwarzschild}}{1958}]{1958RA......5..204S}
{Schwarzschild} M.,  1958, Ricerche Astronomiche, \href
  {http://adsabs.harvard.edu/abs/1958RA......5..204S} {5, 204}

\bibitem[\protect\citeauthoryear{{Thun}, {Kuiper}, {Schmidt}  \& {Kley}}{{Thun}
  et~al.}{2016}]{2016A&A...589A..10T}
{Thun} D.,  {Kuiper} R.,  {Schmidt} F.,   {Kley} W.,  2016, \mn@doi [\aap]
  {10.1051/0004-6361/201527629}, \href
  {https://ui.adsabs.harvard.edu/abs/2016A&A...589A..10T} {589, A10}

\bibitem[\protect\citeauthoryear{{Velasco Romero} \& {Masset}}{{Velasco Romero}
  \& {Masset}}{2019}]{2019MNRAS.483.4383V}
{Velasco Romero} D.~A.,  {Masset} F.~S.,  2019, \mn@doi [\mnras]
  {10.1093/mnras/sty3382}, \href
  {http://adsabs.harvard.edu/abs/2019MNRAS.483.4383V} {483, 4383}

\makeatother
\end{thebibliography}




\bsp	
\label{lastpage}
\end{document}